Milan M. Ćirković

## ANTHROPIC FLUCTUATIONS VS. WEAK ANTHROPIC PRINCIPLE

**Abstract.** A modern assessment of the classical Boltzmann-Schuetz argument for large-scale entropy fluctuations as the origin of our observable cosmological domain is given. The emphasis is put on the central implication of this picture which flatly contradicts the weak anthropic principle as an epistemological statement about the universe. Therefore, to associate this picture with the anthropic principle as it is usually done is unwarranted. In particular, Feynman's criticism of the anthropic principle based on the entropy-fluctuation picture is a product of this semantic confusion.

Various anthropic principles have been discussed during the last quarter of century by cosmologists and philosophers alike (for the bibliography up to 1991, see Balashov 1991; for the most comprehensive treatment of issues involved, see Barrow and Tipler 1986). In spite of this—by now already rather long—history (especially taking into account earlier important contributions like those of Barnes, Whitrow or Hoyle), there are still more than a few misunderstandings and confusing issues in the field of anthropic reasoning. In order to reduce large ambiguities and confusion still reigning in these discussions, here we investigate at some length one of the most important problems allegedly amenable to an "anthropic" solution, namely, the problem of the thermodynamical arrow of time. As we shall see, this is a nice example of actual misuse of the appeal of anthropic principles.

Let us imagine a static, spatially finite and eternally existing universe. For a working model, we can consider classical Einstein (1917) model which marked the beginning of the modern cosmology. This model is a static, topologically closed universe with a cosmological constant. Let us also, for the time being, forget about gravitational collapse, existence of black holes and possible singularities within their horizons, and assume that all interactions are classical, and that the equation of state of matter is always such as to prevent formation of singularities. Now we wish to consider fate of a much smaller region within such a universe, a region which we shall call—for obvious reasons—the observable region,. What conclusions can one draw from the existence of a large thermodynamical disequilibrium in the observable region?

This issue has, historically, been the central part of Boltzmann's thinking about the nature of the second law of thermodynamics. Although he knew nothing about the cosmic microwave background, he knew very well the Olbers' paradox and understood that the thermodynamical disequilibrium is a necessary (although presumably not sufficient) condition for creation of life and ultimately intelligence. Therefore, he suggested two possible recourses: one which postulated very special initial conditions (that is, a universe of finite age, and low initial entropy), and another, that what we see is a large enough fluctuation of entropy creating the local condition of thermodynamical disequilibrium, while all matter is approaching equilibrium reigning in the universe at large. Since any intelligent observers could exist in the disequilibrium condition, it is only reasonable that we perceive the universe as we do, far from equilibrium, and possessing a definite thermodynamical arrow of increasing entropy. In the words of Boltzmann (1895) himself:

> If we assume the universe great enough, we can make the probability of one relatively small part being in any given state (however far from the state of thermal equilibrium), as great as we please. We can also make the probability great that, though the whole universe is in thermal equilibrium, our world is in its present state. It may be said that the world is so far from thermal equilibrium that we cannot imagine the improbability of such a state. But can we imagine, on the other side, how small a part of the whole universe this world is? Assuming the universe great enough, the probability that such a small part of it as our world should be in its present state, is no longer small.

However, this is not stated by Boltzmann as the primary option, and he even gave credit for this idea to his assistant, Dr. Schuetz. The other possibility—listed as the primary option—is the existence of special initial conditions of very low entropy, from which the universe evolves toward states of higher and higher disorder. Now, there exists a problem in an ever-existing universe that the initial conditions need to be formulated in asymptotic limit $t \rightarrow -\infty$, and these difficulties are circumvented only by additional postulates, like that of spacetime as an infinite reservoir of negative entropy (Davies 1974). In evolving universes of finite age, like the Friedmann models, this is seemingly alleviated by postulating a singular origin at $t = 0$, although another problem appears in place of the previous one: improbability of so low entropy Big Bang when compared to the generic case of global singularity (Penrose 1979). In spite of these circumstances, as Barrow and Tipler point out in their (1986) monograph, invoking some kind of special initial conditions was in general much more popular among cosmologists, and for a good reason. Since Hubble's discovery of the expanding universe—and in particular since the victory of the standard Big Bang cosmology over its great steady state rival (Kragh 1996)—the special character of the initial conditions became a part of the core of the standard cosmological paradigm. However, from time to time, the second option has also been put forward, mainly by philosophers, and it is there that the main obstacle to its operationalization has been noted. The intrinsic difficulty in the entropy-fluctuation picture is connected with the very basis of statistical reasoning employed. In words of Price (1996):

> If we wish to accept that our own region is the product of "natural" evolution from a state of even lower entropy, therefore, we seem bound to accept that our region is far more improbable than it needs to be, given its present entropy... If the choice is between (1) fluctuations which create the very low-entropy conitions from which we take our world to have evolved, and (2) fluctuations which simply create it from scratch with its current macroscopic configuration, then choice (2) is overwhelmingly more probable. Why? Simply *by definition*, once entropy is defined in terms of probabilities of microstates for given macrostates. So the most plausible hypothesis—overwhelmingly so—is that the historical evidence we take to support the former view is simply misleading, having itself been produced by the random fluctuation which produced our world in something very close to its current condition.

This is the crux of the problem: our observations in the entropy-fluctuation picture do not correspond to reality—and not for epistemological, but for physical reasons. The argument has been put forward for the first time by von Weizsäcker (1939). The entire evidence pertaining to what is conventionally called past is false or simulated; notably, this applies to cosmological knowledge on the physical state of the universe at previous epochs. Beyond certain region which surrounds our spatiotemporal location, the universe is in thermodynamical equilibrium, but we have no real information on matter in that external region. In the internal region, however, it is only natural to see things in such a way that states of smaller and smaller entropy are envisaged as we look deeper and deeper into the (conventionally labelled) past. If one could apply Boltzmann-Schuetz model to the modern cosmology (which can be, in general, fairly well approximated by Newtonian models, except very close to the initial singularity), this "simulation" would include such paramount observations in modern cosmology, as is the existence and properties of cosmic microwave background radiation or young galaxies seen in modern deep-field images.

However, in order to see the main point of the present study, it is not necessary to track the details of possible application of this model to conventional cosmological wisdom. There is an entirely different aspect of the story, which is seen in Barrow and Tipler's dubbing the Boltzmann-Schuetz idea "anthropic fluctuation picture", therefore implying its close connection with at least some of the various anthropic principles. The present author has argued elsewhere that at least some of the so-called anthropic principles are certainly not principles, but workable scientific hypotheses (Ćirković and Bostrom 2000). What exactly is "anthropic" in the Boltzmann-Schuetz entropy fluctuation model?

The central idea is, of course, that we could not exist in the state of thermodynamical equilibrium, and that, therefore, some restrictions on *possible* worlds are imposed by our existence as intelligent observers in this specific case. Is that enough to constitute anthropic nature of an argument? This does not seem obvious at all. For instance, nobody has invoked anthropic reasoning when discussing incapability of other planets in the Solar system to support life and intelligence; underlying physical mechanisms in this case are (allegedly) sufficiently well understood (e.g. Dole 1964) that the teleological mode of explanation can be substituted by a more conventional one.[1] *Weak anthropic principle* (henceforth WAP), the very foundation of entire anthropic reasoning, has been originally defined by the following locution (Carter 1974):

> …the effect that we must be prepared to take account of the fact that our location in the universe is *necessarily* privileged to the extent of being compatible with our existence as observers. (p. 127)

Alternatively, one can use the definition of Barrow and Tipler (1986):

---

[1] In this connection see the analysis of explanatory power of anthropic arguments by Balashov (1990); somewhat different opinions are expressed in Earman (1987) and Kragh (1997).

The observed values of all physical and cosmological quantities are not equally probable but they take on values restricted by the requirement that there exist sites where carbon-based life can evolve and by the requirement that the Universe be old enough for it to have already done so. (p. 16)

Tha latter does explicitly deal with *observations*, and not "underlying reality" (or just "values of all… quantities"). However, as a statement on observations, it possesses no true epistemological value. The reason for this (seemingly too strong) thesis can be found in subsequent discourse of the same authors in discussion, for instance, of the dimensionality of space, one of the great questions of theoretical physics which is an obvious field for anthropic reasoning. The very title of the seminal Ehrenfest (1917) article on the topic (discussed at length in the Barrow and Tipler monograph, pp. 258-276) speaks for itself: *In what way does it become manifest in the fundamental laws of physics that space has three dimensions?* The usage of "manifest" instead of, say, "apparent" or "perceived" indicates that the explanation is to be offered through anthropic reasoning for something taken at the phenomenal level. This should be compared with "simulations" of reality implicit in the Boltzmann-Schuetz picture; while those can be interpreted as "observed values" in Barrow and Tipler definition, there is contradiction with the Carter's requirement that *our location* (on which data are *not* simulated) imposes true restrictions on the state of the universe at large. This seems another instance in which Barrow-Tipler definition is inferior to Carter's, from a physical—one may even say practical—point of view (for more on the issue of definitional differences see Bostrom 2002).

In the amount in which WAP is a cornerstone of the entire anthropic reasoning, Boltzmann-Schuetz idea is, therefore, antithetical to the substance of the latter. It implies not only a negation of the cognitive value of our observations, but also breaks the basic connection between the local and global processes. The point of the WAP constraints, as there were conceived by Dicke and explicated by Carter, has been exactly to deal with real properties of the universe, not how the universe could or should look like. It should not just *look* flat enough; it *must* really be flat in order for WAP to meaningfully work. Presumably, a thesis suggesting that WAP is dealing only with observations of ours with all their inherent limitations can be defended.

However, in the view we are defending here, it would present a betrayal of the spirit of the entire anthropic reasoning, and—what is much worse—would ultimately require a revision of most of our basic assumptions first of all in cosmology and particle physics, and contingently in other scientific disciplines as well. Obviously (from the context of their writings) that Dicke, Carter, Barrow and Tipler, and other authors, have not been considering the possibility that our empirical evidence is just a mirage.

This is, of course, not to imply that the entropy-fluctuation picture is simply wrong, even in the framework of our classical, non-expanding universe. However, there is a host of epistemological problems with it which make it at least highly suspicious. Whether a theory implying simulated nature of most empirical evidence may be considered scientifical at all is not obvious. For instance, there seems to appear problems with falsifiability in the Popperian sense. As far as cosmology is concerned, in a founding paper of the historically crucial steady state theory, Bondi and Gold (1948) have emphasized the desirability of as uniform physical laws as possible. In addition, the conceptual ease of empirical refuting the steady state theory made it a paradigm of the Popperian notion of scientific theory, as later elaborated by Bondi (1967). Although the steady state theory is now universally considered defunct, its impact in the epistemology and philosophy of cosmology has been instrumental in the formation of modern cosmological thought. In the now accepted standard paradigm based on Friedmann models the uniformity of laws is preserved in entire spacetime except at initial (and possibly final) singularity. The Boltzmann-Schuetz picture is an extreme example of opposite attitude, in which the very concept of physical law becomes vacuous. Therefore, it is hardly a scientific theory in Popperian sense. Parenthetically, the entropy-fluctuation picture for the same reasons violates Reichenbach's principle of the common cause, which states that improbable coincidences are always associated with an earlier common cause. Therefore, it puts us in an uncomfortable position of having to consider only sufficiently isolated systems as causally ordered. This potentially undermines most of usual notions of description in physical sciences, since one can never be entirely certain that his system is isolated enough. A similar (but less disturbing) situation occurs in some other attempts to make local physics dependent on largely unknown and potentially unknowable cosmological processes or boundary conditions (for instance, in Hoyle and Narlikar attempts at building an action-at-distance description of inertial properties of matter; e. g. Hoyle and Narlikar 1972; Hoyle 1975).

One can reach the same conclusion from another point of view. The entropy fluctuation picture (correctly) and the anthropic principle itself (in our opinion incorrectly) have been criticized on the basis of what can be called "Adam's dilemma" (from Milton's *Paradise Lost*); the most widely cited formulation of this criticism is due to Feynman (1965). The great physicist concluded that the entropy fluctuation picture is "ridiculous" since it requires much larger entropy fluctuation in both spatial and temporal sense than the one necessary for the actual emergence of intelligent observers; thus it is in conflict with Occam's razor. This criticism, while *prima facie* irreproachable, applies seriously only to the classical, Boltzmann discussion. As we have noted, Boltzmann and Schuetz have actually known next to nothing about what we perceive today as the state of the universe on cosmological scales, and could rely only on a very limited set of local (in spatiotemporal sense) observations. Moreover, even today it is not at all obvious *how large* entropy fluctuation is *actually* needed for the emergence of life and intelligence on Earth. As an antidote to Feynman's cognitive optimism, one may cite the classical study of Collins and Hawking (1973), where it was argued that properties of spacetime on the *largest* scales are necessary preconditions for our existence (the detailed discussion of this issue is to be found in Barrow and Tipler 1986). Boltzmann, however, was quite correct that *any size* of fluctuation can, ultimately, be achieved in his globally static world of classical physics; in this sense, Feynman may only criticize this picture on account of its *improbability*, i.e. sufficiently small measure on a subset of all conceivable observer-creating fluctuations. This measure is small indeed, at least in the spatial sense, since the observed universe is so much bigger than the terrestrial biosphere (as far as the temporal scale is concerned, the case is much more difficult to establish, since the timescales for evolution of intelligence are comparable to cosmological timescales in realistic Friedmann-Robertson-Walker universes). However, this problem is entirely circumvented by recalling the overwhelming probability that our empirical knowledge is simulated in this picture. The fluctuation could indeed be much smaller, as Milton's Adam, Feynman and others supposed; however, we do not need to know about this. In fact, it is due only to the widespread confusion plaguing the field that Feynman has reportedly criticized *anthropic* *principle*, when discussing the entropy-fluctuation picture. His criticism is in place when applied to the conventional form of the latter, but the picture itself has nothing to do with the anthropic principle whatsoever. Interestingly enough, in the course of this criticism, Feynman has put forward a necessity for a time-asymmetric law of nature to explain the entropy gradient, a proposal which was later strongly publicized by Penrose (for instance see Penrose 1979).

It has been established long ago, notably by great French physicist, mathematician and philosopher Henry Poincaré (1946)—and later elaborated by the founder of the modern computer science Wiener (1961)—that the sense of increasing entropy is essential for existence of intelligent observers. Without entering into discussion on whether other forms of complex organization of matter can exist in an entropy-decreasing universe, one notes that in the Boltzmann entropy fluctuation picture, there is necessarily a finite period of time in which the entropy did actually decrease. This period, therefore, necessarily *preceeded* the appearance of intelligent life on our planet. Together with our knowledge about the history of human intelligent observers, this gives us a lower limit to our local displacement relative to the minimum of the entropy curve. However, since one can envisage the time when exact mechanism of evolution and processes accompanying emergence of intelligence will become known facts, it seems *a posteriori* improbable that a smooth transition between "simulated" (cosmological) and "real" (anthropo-biological) evidence could be achieved. This may be regarded as a possible (still distant, of course) argument against the entropy fluctuation picture *per se*, although we can not devote more attention to this issue here.

The present discourse applies to the classical universe, considered as a single topologically connected cosmological domain. In the recent literature in both cosmology and philosophy, there is a surge of interest in the concept of *multiverse*—set of different cosmological domains, possibly causally and/or topologically disconnected from our "observable" domain—as the most comprehensive description of everything that exists. A variation on the Boltzmann-Schuetz theme can be played within the multiverse framework, in which our existence as observers selects a particular domain ("universe"), as a domain of exceptionally low entropy content. A detailed discussion of this idea is beyond the scope of the present work, but two points may be made immediately. Since the anthropic selection effect affects initial conditions ("Big Bangs") of multiple universes, such a proposal would arguably be closer to the Boltzmann's primary option than to the entropy-fluctuation picture. On the other hand, such an approach would essentially vindicate our observational knowledge on the structure of our domain on the largest scales, thus leading to no epistemological quandary of the sort discussed in the present article.

In conclusion, modern understanding of the entropy-fluctuation picture immediately implies the simulated nature of most of our cosmological evidence (as well as all other evidence pertaining to the perceived past). This does not correspond to the phenomena in common empirical sense of the word, and therefore immediately implies a massive violation of the WAP constraints. Thus, the attribute "anthropic", arbitrarily associated with this picture, is largely shallow and superficial. By unravelling this semantical confusion, we immediately gain a proper account of the problem von Weizsäcker, Feynman and others have perceived, which can not truly jeopardize the epistemological and methodological status of WAP (and, contingently, other anthropic principles incorporating WAP as their core).

**Acknowledgements**. The author wishes to hereby thank to Vesna Milošević-Zdjelar for supplying some of the references, as well as to Dr Nick Bostrom and Prof. Petar Grujić for stimulating discussions.

*Astronomical Observatory of Belgrade*
*Volgina 7, 11160 Belgrade, Serbia, YUGOSLAVIA*